\documentclass[traditabstract]{aa} 
\usepackage{graphicx}
\usepackage{aalongtable}
\usepackage{txfonts}
\usepackage{natbib}
\usepackage{rotating}
\usepackage{lscape}
\newcommand{\gsim}{\;\lower.6ex\hbox{$\sim$}\kern-7.75pt\raise.65ex\hbox{$>$}\;}
\newcommand{\lsim}{\;\lower.6ex\hbox{$\sim$}\kern-7.75pt\raise.65ex\hbox{$<$}\;}

\newcommand{\ebv}{E$(B-V)$}

\newcommand{\y}{$y_0$}
\newcommand{\cy}{$c_{y,0}$}
\newcommand{\m}{$m_0$}
\newcommand{\del}{$\delta_4$}

\begin{document}
\title{A Str\"omgren view of the multiple populations in globular clusters 
}

\author{
Eugenio Carretta\inst{1}
\and
Angela Bragaglia\inst{1}
\and
Raffaele Gratton\inst{2}
\and
Valentina D'Orazi\inst{2}
\and
Sara Lucatello\inst{2}
}

\authorrunning{E. Carretta et al.}
\titlerunning{Str\"omgren photometry and MPs in GCs}

\institute{
INAF-Osservatorio Astronomico di Bologna, Via Ranzani 1, I-40127
 Bologna, Italy\\
\email{eugenio.carretta@oabo.inaf.it, angela.bragaglia@oabo.inaf.it}
\and
INAF-Osservatorio Astronomico di Padova, Vicolo dell'Osservatorio 5, I-35122
 Padova, Italy\\
 \email{raffaele.gratton@oapd.inaf.it, valentina.dorazi@oapd.inaf.it, sara.lucatello@oapd.inaf.it}
  }

\date{}

\abstract{ We discuss a variety of photometric indices assembled from the
$uvby$  Str\"omgren system. Our aim is to examine the pros and cons of the
various indices to find the most suitable one(s) to study the properties of
multiple populations in globular clusters (GCs) discovered by spectroscopy.  We
explore in particular the capabilities of indices like $m_1$ and $c_y$ at
different metallicities. We define a new index $\delta_4=(u-v)-(b-y)$  to
separate first and second stellar generations in GCs of any metal abundance,
since it keeps the sensitivity to multiple stellar populations over all the
metallicity range and at the same time minimizes the sensitivity to photometric
errors. We detecte clear differences in the red giant branches of the GCs
examined, like skewness or bi/multi-modality in color distribution. We connect
the photometric information with the spectroscopic results on O,  Na abundances
we obtained in our survey of GCs. Finally, we compute the effects  of different
chemical composition on the Str\"omgren filters and indices using  synthetic
spectra. }  

\maketitle

\section{Introduction}\label{intro}

\begin{figure*} 
\centering 
\includegraphics[]{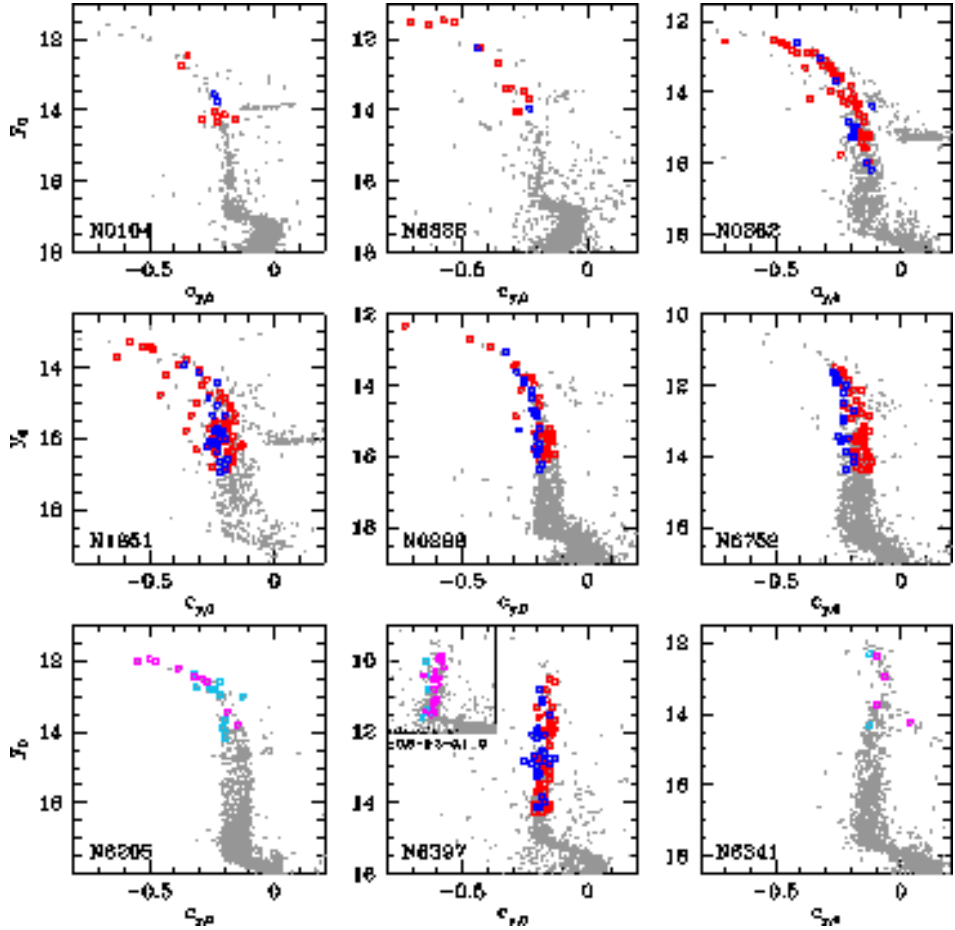}
\caption{CMDs in \y,\cy \ for the nine GCs, seven of which in common with our
spectroscopic survey. Blue symbols indicate P stars and red symbols IE stars,
all confirmed members on the basis of their RVs.  For NGC~6205 (=M13) and NGC~6341 (=M92), whose 
spectroscopic data are not homogeneous to the seven other GCs, we use
light blue and magenta, respectively; there are two suspect AGB stars in 
NGC~6205 and one confirmed HB star in NGC~6341 (note that the HB is redder than the 
RGB in these plots). For NGC~6397 we display in an inlet
the results from \cite{lind09,lind}, shown on our same scale in 
magnitude, and on an expanded scale in \cy; the separation between 
Na-poor (light blue) and Na-rich (magenta) stars
is perhaps cleaner, but is based on a smaller sample (about 30 stars).
}
\label{figroma2} 
\end{figure*}

\begin{figure*}
\centering
\includegraphics[scale=0.9]{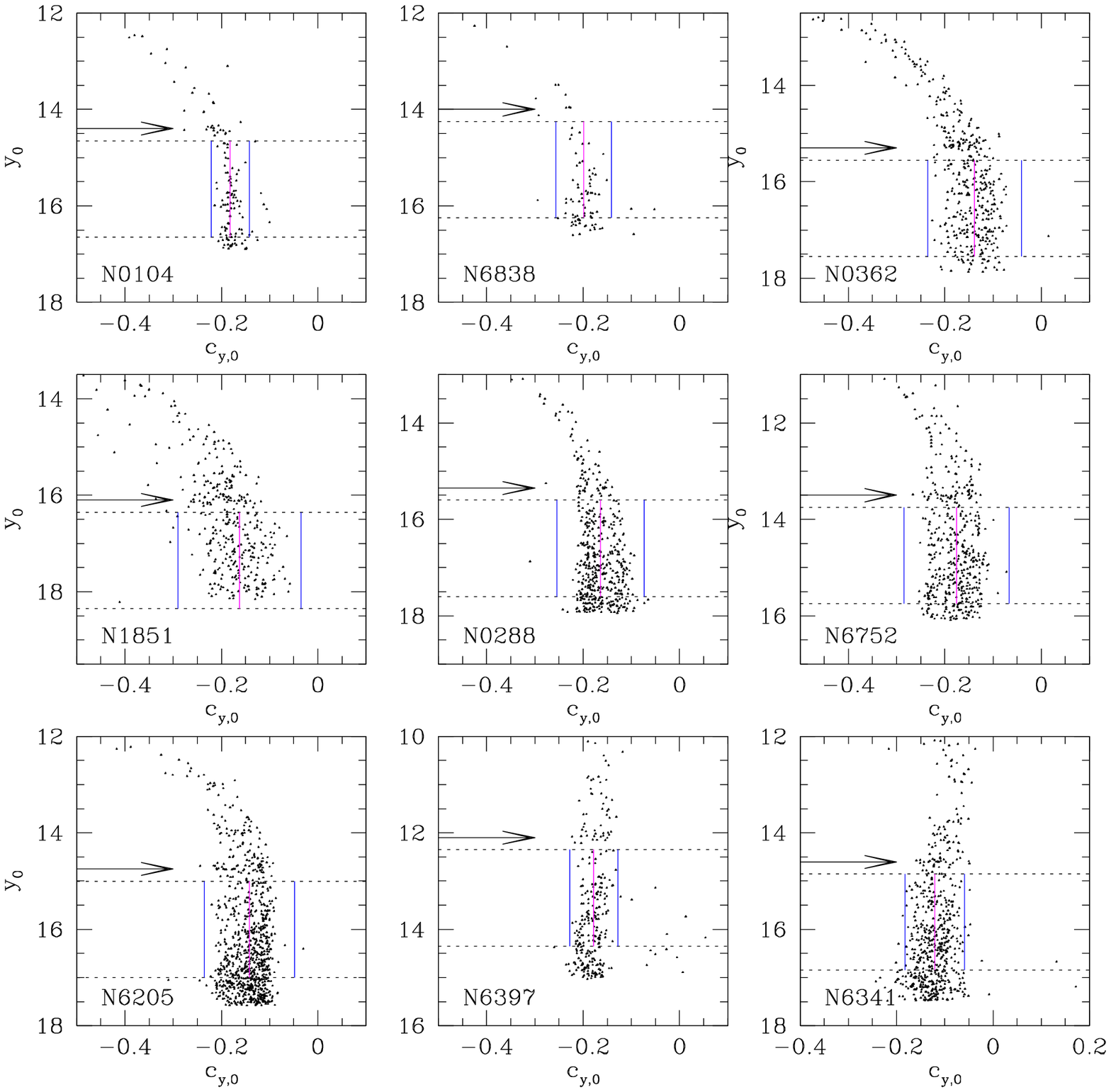}
\caption{Enlargement of the RGB region for the CMDs in \y,\cy \ of the 
nine GCs. An arrow shows the assumed magnitude of the RGB bump. The 
lines indicate the RGB stars selected to derive average and r.m.s. values 
for \cy \ (see Sect.~\ref{param}).}
\label{cymed}
\end{figure*}

Thanks to the advent of efficient multi-object spectrographs at 8\,m-class 
telescopes and consequent extensive surveys, in the last few years it has 
become clear that the presence of multiple stellar populations is ubiquitous in
globular clusters (GCs) \citep[see,
e.g.,][]{marinom4,carretta09a,carretta09b,carrettam54lett,carretta1851a,
carretta1851b,cj}. There are at least two generations of stars, the second one
formed from the ejecta of a fraction of the first generation (the so-called
polluters, see e.g., \citealt{gratton01,araa}). This is a crucial piece of
information on the early phases of evolution of GCs. It seems to be related to
their intrinsic, specific formation mechanism  \citep{carretta06} and may
give constrains to theoretical models of formation of  massive stellar
clusters. 

With some delay with respect to spectroscopy, photometry is now also indicating
the presence of different stellar populations (not necessarily of different
age)  in a subsample of clusters \citep[see e.g.][]{anderson98,bedin04}. We will discuss
some of its potentialities in the present paper, including strong sensitivity to
abundances of some crucial element (namely N) and availability of large samples.
However,  most of the ``hard" evidence concerning multiple populations in GCs is
still  based on spectroscopic surveys, where the multiple populations manifest
through a  distribution of stars along the Na-O, Mg-Al, and C-N
anti-correlations, that  have been found so far in $all$\ GCs where abundances
of these elements in a large enough number of stars  have been measured.
Typically,  modern spectroscopic surveys devoted to these studies include at
best about 100 stars per  cluster \cite[e.g.][and references
therein]{carretta09a}  with internal errors in abundance, derived from
intermediate and high resolution spectra, typically representing about 8-10\% of
the total variance.  While  various general information about early evolution of
GCs can be obtained from such distributions (e.g. the number ratios between
first and  second generations and the total extent of the Na-O anticorrelation,
see Carretta et al. 2009a, Carretta et al. 2010c), other require both larger
samples and smaller internal errors. In particular, this is the case for the
distinction between smooth continuous distributions and discrete ones. Such a
distinction is crucial, because it provides a basic clue for the nature of the
polluters, that might be either fast rotating massive stars  \citep[lifetime a
few $10^6$~yrs:][]{decressin} or the most massive among intermediate mass stars,
undergoing hot bottom burning during their AGB phase  \citep[lifetime a few
$10^7$~yrs:][]{ventura01}. For instance, discrete distributions might be only
compatible with  the AGB polluters, as suggested by \cite{renzini}. 

Moreover, while multiple populations are an intrinsic property of $any$ 
GC \citep[so that this might be adopted as definition of a genuine GC, 
see][]{carretta10c}, the existing statistics shows also that $each$
GC does show difference in the shape and extensions of the Na-O
and Mg-Al anticorrelations. In other words, apart from a common nucleosynthetic
pattern, each cluster is characterized by its particular history, whose
features may be studied, in principle, with high resolution spectroscopy of
a very large number of stars.

Alternative to spectroscopy, and less time consuming, photometry can provide
both large samples  and smaller internal errors. The most suitable photometric
systems should couple a  large sensitivity to the abundance variations related
to the multiple population phenomena, with a good enough efficiency and the
possibility to be applied to wide-field photometry. Sensitivity is obtained by 
passbands including strong molecular bands of CN (bandheads at 3883 and
4216~\AA) or  NH (around 3450~\AA) or CH (around 4300 \AA), since there are no
common use filters sensible to  the (smaller) spectral variations due e.g., to
different Na abundances.  Several photometric systems have been considered,
including broad band (see e.g. \citealt{carretta10d,lardo}), 
intermediate band \citep[see][]{gru99}, and narrow band ones  \citep{lee09}.  
these different methods have advantages and limitations: for instance
Sloan filters \citep{stoughton} are very efficient, and high quality photometry can be obtained
even for faint sources. However, sensitivity to the abundance variations is
limited. On the other hand, while narrow band photometry may have very high
sensitivity, results are not well understood yet (see Lee et al. 2009, and
Carretta et al. 2010e). In this paper we will consider in more detail results
from Str\"omgren photometry, extending the previous work by \cite{gru99} and  \cite{yong08}.

Str\"omgren photometry \citep{strom56,strom66} has been originally devised for 
F-dwarfs and is
applied mostly to the study of hot and intermediate spectral type (O to G) 
main sequence stars. There is however a long series of studies,
\cite[see e.g.][]{grundahl98,gru99,gru02,att00,attm07} showing that it can be successfully used also for 
stars on the red giant branch (RGB) and on the horizontal branch (HB). 
In particular, \cite{att00} briefly discussed the 
effects of the "abundance anomalies" on the photometry of RGB stars in GCs.
\cite{grundahl98} presented an extended Str\"omgren
photometry of NGC6205 (=M13) and noted how the RGB stars presented a large spread in the
$c_1$ index. They attributed it to the
different chemical composition (not different iron content) of stars, since the
relation between $c_1$ and CN, CH  strength had already been demonstrated
\citep[e.g. in M~22 by][]{attc95}.
A direct  measure of N abundance was obtained by \cite{yong08} and was used
to calibrate a new index $c_y$ (see Sect.~\ref{cy}). 
A similar approach was followed for instance by \cite{carretta09a} that
connected the $c_y$ index to stars of the first (Na-poor) and of the second 
(Na-rich) generation in NGC~6752.

\cite{sbordone} have recently presented a theoretical study of the influence 
of non-standard composition (e.g., enhanced in Na and/or He, and/or light 
elements like CNO) on the stellar models and the resulting isochrones, having 
in mind the particular case of the multipopulations in GCs. They considered 
the cases of Johnson $UBVI$ and Str\"omgren $uvby$ filters. We will discuss 
later (Sect.~\ref{sbordone}) the similarities and differences between their 
completely theoretical work and our observational approach. However, we note 
immediately that one of the most important is that they did all computations 
for a single metallicity\footnote{Within this paper, the term metallicity
is usually adopted as a synonym of [Fe/H], with an overall abundance pattern
characteristics of halo stars, that is with an excess of $\alpha-$elements.}.
 value, near the average for the Galactic GC
population,  while we did instead consider GCs spanning (almost) the entire
metallicity  range for Milky Way GCs, from about $-2.25$ to about $-0.70$
\,dex. 

In this paper we will systematically and deliberately  try to determine whether 
photometric data, in particular in Str\"omgren filters, can be used to
discriminate  between stars of different stellar generations in GCs, by looking
at their RGBs. We see that the situation is rather complex, and that there is a
strong  dependence on the evolutionary phase and the metallicity considered. For
instance, the index $c_y$ is efficient in  separating different  populations in
GCs, but only at intermediate/low metallicity, while it loses sensitivity for
metal-rich clusters (at least for the lower RGB).  This sensitivity might even
change its sign, making the interpretation of observational data ambiguous. The
$m_1$ index is instead a good indicator of multiple populations at high
metallicity. Finally, we explore the various indices that is possible to define
and we select a new index $\delta_4$ that uses all the four Str\"omgren filters
$uvby$.  We show that it is almost independent from reddening and that it can be
used  to efficiently separate different stellar generations in GCs over a large
range in metal abundance.

The plan of the paper is as follows: in Sect. 2 we present the available data on
which we base our analysis and we discuss the separation of GC stars in  first
and second generations as seen by the classical indices already known:  $c_y$
and $m_1$. In Sect. 3 we examine other possibilities and present our new index
$\delta_4$, exploring its sensitivity to metallicity and to the properties of
different stellar generations. In Sect. 4 we study the correlations  between \cy
\ and \del \ and cluster parameters. In Sect. 5 we interpret our results with
the aid of synthetic spectra. A summary is given in Sect. 6.

\begin{figure} \centering 
\includegraphics[]{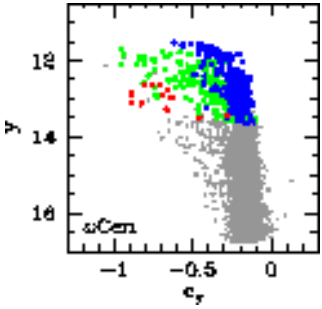}
\caption{
CMD (not corrected for reddening) of $\omega$ Cen with stars of different
[Fe/H] (from Johnson \& Pilachowski 2010) indicated by different colors:
blue stars have [Fe/H]$<-1.6$, red ones $>-1$, and green ones are in between.
}
\label{omega} 
\end{figure}

\begin{figure*}
\centering
\includegraphics[bb=30 520 580 695, clip, scale=0.95]{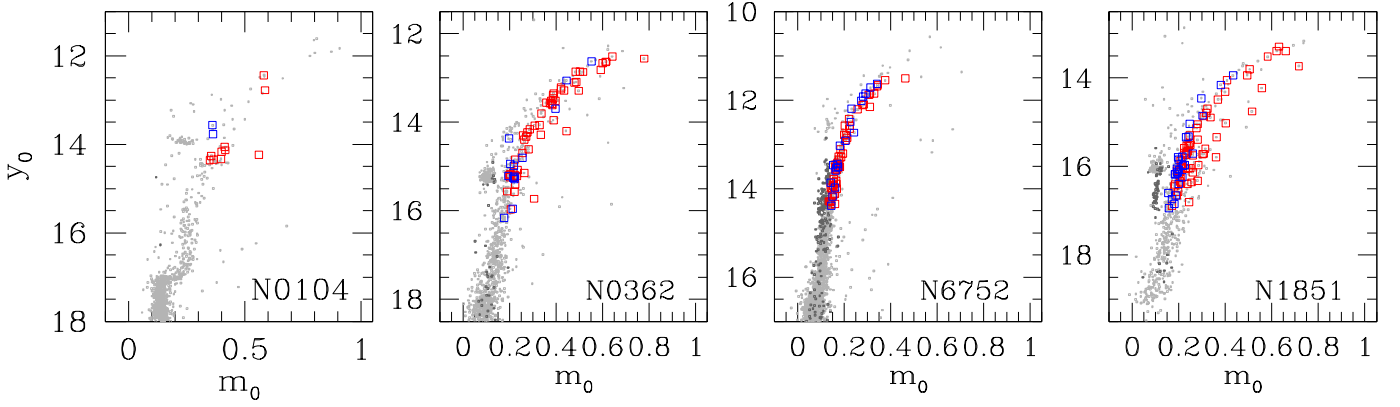}
\caption{CMDs in \m, \y \ for three monometallic GCs and NGC~1851.  
Blue and red symbols have the same meaning of Fig.~\ref{figroma2}; 
darker grey points, visible for NGC~6752 and NGC~1851, are 
blue HB stars (as indicated by their $b-y$ color).
}
\label{figroma3}
\end{figure*}

\section{The data}\label{sec2}

We used the publicly available Str\"omgren photometry for nine Galactic GCs
collected by Grundahl and coworkers \citep{grundahl98,gru99, gru02}
and  presented by \cite{calamida}, who used them to obtain  a new calibration
of  the metallicity index $m_1$\footnote{We downloaded the catalogues from the
web page {\tt http://www.oa-roma.inaf.it/spress/gclusters.html}}.   These data
were obtained at  two different telescopes (Nordic Optical Telescope --NOT-- on
Canary Islands, Spain, and  Danish telescope on La Silla, Chile), as indicated
in Table~\ref{data}. Instruments  covering very different fields of view (about
4\arcmin \ and 11\arcmin ~on a side) were used; sometimes  a single field was
targeted, sometimes two or more. The observations generally comprise  the GC
center, with the exception of the two most metal-rich cluster in the sample,
NGC~104 (=47 Tuc) and  NGC~6838 (=M71). We refer to the \cite{calamida}  paper for references 
and details on the observations and calibrations.

We made a selection on photometric errors and sharpness (a parameter used to
separate well measured stars from galaxies, defects, and cosmic rays), retaining
only those stars  with errors in all the four filters $\le 0.02$ mag and 
$|{\rm sharpness}| \le 0.2$.  We used the 2MASS Point Source Catalog 
\citep{2MASS} to
astrometrize the catalogues through software written by P. Montegriffo at the
Bologna Observatory. We employed the \ebv \ values from \cite{harris} and
corrected the $u,v,b,y$ magnitudes  for reddening using the following relations 
$ u_0=u-5.231\times E(B-V),$
$ v_0=v-4.552\times E(B-V),$
$ b_0=b-4.049\times E(B-V),$
$ y_0=y-3.277\times E(B-V),$
with coefficients taken from Tab.~6 of \cite{sfd98}. From now on, all
the quantities,  colors, and combination of bands are intended as dereddened.
For seven of the clusters  the reddening is very low (\ebv=0.02 to 0.05). It is
slightly higher for NGC~6397, but  only for NGC~6838 some differential reddening
is present. Finally, we estimated the  magnitude of the RGB bump visually
inspecting the color-magnitude diagrams (CMDs); no  sophisticated technique was
applied, since this information is used only to roughly  guide our selections
(see below).

\begin{table}
\centering
\caption{List of GCs.}
\setlength{\tabcolsep}{1.5mm}
\begin{tabular}{llcccl}
\hline
~~~~~~GC       &Other  &[Fe/H]&\ebv &\y(RGB$_{bump}$) &Telescope\\
\hline
\object{NGC~~~104} &47 Tuc &-0.76 &0.04 &14.40 & Danish\\
\object{NGC~6838}  &M71    &-0.82 &0.25 &14.00 & NOT\\
\object{NGC~~~362} &	   &-1.17 &0.05 &15.30 & Danish\\
\object{NGC~1851}  &	   &-1.18 &0.02 &16.10 & Danish\\
\object{NGC~~~288} &	   &-1.32 &0.03 &15.35 & Danish\\
\object{NGC~6752}  &	   &-1.55 &0.04 &13.50 & Danish\\
\object{NGC~6205}  &M13    &-1.58 &0.02 &14.75 & NOT\\
\object{NGC~6397}  &	   &-1.99 &0.18 &12.10 & Danish\\
\object{NGC~6341}  &M92    &-2.25 &0.02 &14.60 & NOT\\
\hline
\end{tabular}
\tablefoot{[Fe/H]: Carretta et al. (2009c) and Carretta et al. in preparation 
for NGC~362; \ebv: Harris (1996); \y(RGB$_{bump}$): this work.}
\label{data}
\end{table}

\begin{figure*}
\centering
\includegraphics[]{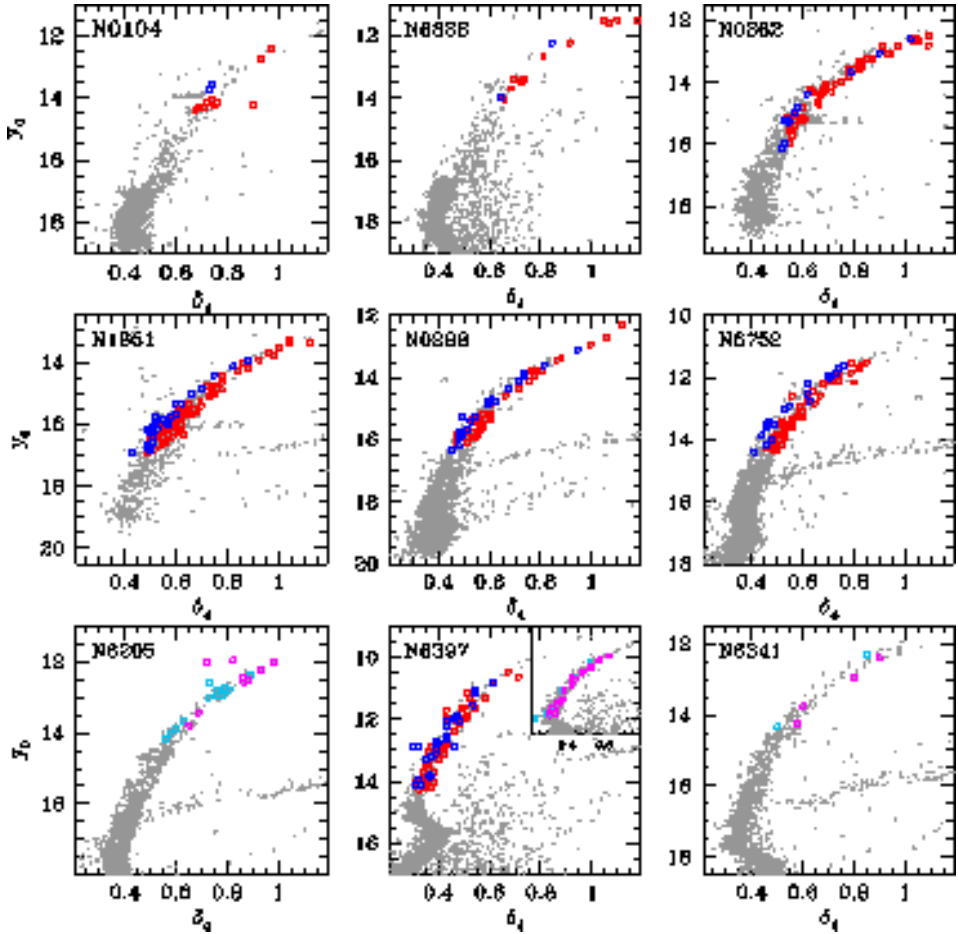}
\caption{CMDs in \del,\y \ (see text for the definition of  $\delta_4$). 
Blue symbols indicate P stars and red symbols IE stars (light blue and 
magenta for NGC 6205 (=M13) and NGC6341 (=M92)). Note that the HB is redder than the RGB in 
these plots for intermediate and low metallicity and that the blue HB 
and the blue stragglers are almost horizontal sequences. As in Fig.~1, for NGC~6397 we also show 
the results from \cite{lind09,lind}; the separation between 
Na-poor (light blue) and Na-rich (magenta) stars
seems cleaner, due to the better data quality and the smaller sample.
}
\label{figroma4}
\end{figure*}

\subsection{P, I, and E stars in Str\"omgren photometry} \label{pie}

For seven of the clusters we obtained FLAMES spectra of RGB stars in our ongoing
survey to study the Na-O anticorrelation in a large sample of GCs with different
global parameters (see Carretta et al. 2007a for NGC~6752;  Carretta et al.
2009a,b for NGC~104 (47~Tuc), NGC~6838 (M~71),  NGC~288, and NGC~6397;  Carretta
et al. 2010b, 2011 for NGC~1851; Carretta et al., in prep. for NGC~362). We 
counter-identified stars in our spectroscopic samples, containing 50-150
objects 
each, and in the Str\"omgren catalogues. We generally found about 100 or more 
stars in common, except for NGC~104 and NGC~6838, where they are about 15.  In
the last two cases this is due to the fact that our FLAMES fiber configurations
are positioned around the cluster center, whereas the  photometric observations
are located off the cluster center. We  use these stars
to try to understand whether it is possible to discriminate  stars of first and
second generation (primordial -P- and intermediate plus extreme -I plus E-
respectively, in our  notation: see Carretta et al. 2009a for the exact
definition) on the basis of their position in color-magnitude diagrams (CMDs),
experimenting  several combination of filters.  All stars analysed
spectroscopically are members of the clusters, based on their derived radial
velocities and chemical abundances.

For NGC~6205 (=M13) and NGC~6341 (=M92) we did a similar identification, using literature 
papers. For NGC~6205, a cluster presenting an extended  Na-O anticorrelation 
\citep{sneden04,cohen05}, we used the measures of Na in \cite{johnson05} 
for the stars in common with our photometry (22 in total, two of which 
seem AGB, not RGB stars) to distinguish P and IE stars, since this paper 
includes more stars than the two cited above.  For M92 there are no 
extensive studies of the Na-O anticorrelations (\citealt{armosky} 
measured Na and O in only nine stars), but 34 stars have Na abundances 
in \cite{sneden00}. We counter identified objects in  Sneden et al. with 
the photometric data and found six stars in common (one of them 
is an HB star).

\begin{table}
\centering
\caption{Str\"omgren indices used in the paper.}
\begin{tabular}{llll}
\hline
Index & Definition & Alternative &Sensitive to:\\
\hline
$m_1$        & $(v-b)-(b-y)$ &$v-2\times b+y$ & metallicity\\
$c_1$        & $(u-v)-(v-b)$ &$u-2\times v+b$ & gravity\\
$c_y$        & $c_1-(b-y)$   &$u-2\times v+y$ & gravity \& N\\
$\delta_4$   & $(u-v)-(b-y)$ &$c_1 + m_1$     & new index\\
$\delta_4$'  & $(\delta_4-\delta_{4B})/(\delta_{4R}-\delta_{4B})$ &		      & no $T_{\rm eff}$  dependence\\
$\delta_4$"  & $\delta_4' \times(\delta_{4R,b}-\delta_{4B,b})$	  &		      &\\
\hline
\end{tabular}
\tablefoot{$\delta_{4B}$ and $\delta_{4R}$ are the \del \ values 
derived using the blue and red polynomial at the $y_0$ of each star 
(see text and Fig.~\ref{delta4all}). $\delta_{4B,b}$ and $\delta_{4R,b}$ 
are the same, but derived at the magnitude of the bump.}
\label{indici}
\end{table}

\subsubsection{The $c_y$ index} \label{cy}

Yong et al. (2008) defined the index $c_y$ (see Table~\ref{indici} 
for a summary of all the usual and new Str\"omgren indices used in 
the present paper) in their paper on NGC~6752; it traces the N 
abundance and was introduced because it is insensible to the 
temperature, at first order, as visible from the 
near-verticality of the RGB at least up to the level of the RGB bump 
(see Fig.~\ref{figroma2}). The dependence  of  $c_y$ to N
abundance comes from its definition, since it contains both the $u$ filter, 
where the strongest NH bands are present (at $\sim 3450$~\AA) and the $v$ one,
where the CN features are present (they are weaker but the filter is weighted
twice in this index). Given the definition of $c_y$, we might then
expect sensitivity to N abundance but also a complex behaviour. In fact,
at low metal abundances and high temperatures only the NH contribution
should be important, CN being negligible due to the combination of the
quadratic dependence on metal abundances (N and C) and the strongest
sensitivity to temperature of its formation. However, at high metallicity,
absorption due to CN is much more relevant, and may dampen or even
change the sign of the sensitivity of $c_y$ on N abundances. 
More details are given in Sect.~\ref{synth}.

Fig.~\ref{figroma2} shows the \y,\cy  \ CMDs for the nine GCs, seven 
of which are in the FLAMES spectroscopic survey. As evident also from 
the enlargement of the RGBs presented in Fig.~\ref{cymed} (where the 
lines will be described in Sect.~\ref{param}), the structure of the RGBs 
is different in the various clusters: the distribution in \cy \ can be 
rather uniform and define a more or less tight sequence (as in NGC~104 (=47 Tuc) or NGC6341 (=M92)), 
or present a wide dispersion (NGC~1851), or show indications of skewness NGC~6205 (=M13)
and bi-modality (NGC~288, NGC~6752). 
In Figure 1 we identify P and IE stars with different colors (blue for P and
red for IE stars, and we maintain the color-coding throughout the paper). The P
stars (i.e., Na  and N-poor first generation stars)   tend to lie on a tight
sequence on the blue side of the RGB, while the I and E ones (i.e., Na and
N-rich stars of the second generation), hereafter considered all  together,
occupy a larger color range on the red side. This is most evident in  NGC~288,
NGC~1851, and NGC~6752, and also clear in NGC~362 and NGC~6397.  Note that in 
\cite{carretta09a} we could classify as P, I, and  E only 16 objects for
NGC~6397, because we conservatively required both Na and O abundances. However,
to  separate in P and IE populations, it is necessary to know only Na, so that
we  have all our 100 stars available for comparison. Only a few
spectroscopically studied stars are present for NGC~104 and NGC~6838, but the
two populations do not separate very clearly.  

Such a segregation in color has already been found by \cite{marinom4} in M4
(using the Johnson  $U$ filter); in NGC~6752 by \cite{carretta09a}, where we 
identified the N-rich stars in \cite{yong08} with the Na-rich ones; and in
NGC~6397 by \cite{lind}, using the Str\"omgren index $c_y$ (and  tightening the 
connection between abundance variations of different elements). The same effect
(and the same explanation) was explored by \cite{lardo} in nine GCs using
the Sloan $u$ filter.

Our first conclusion is therefore that \cy\  {\it seems to efficiently
separate sequences of first and second generation stars at intermediate  
to low metallicities (indicatively for [Fe/H]$\lsim-1.3$), but does not 
work well for  metal-rich clusters like NGC~104 or NGC~6838, or even NGC~362}. 
We will see in Sect.~\ref{synth} that the effects of chemical abundances 
are complex.

In NGC~1851, even if the two stellar populations, P and IE, are
segregated in \cy\ along the RGB for $y_0>15.5$, the branch has a larger 
spread in Fig.~\ref{figroma2} with respect to the other GCs. Recently 
\cite{carretta1851a} presented  the hypothesis that NGC~1851 is the result 
of a merger of two originally distinct GCs with slightly different
metallicity, each one with a Na-O anticorrelation. The appearance of the plot
would be explained if we are  seeing the effect of a (small) spread in 
[Fe/H] and likely overall CNO content \citep[]{yg08}, with Na-O 
(hence N) differences in both the metal-rich and metal-poor components. 
That $c_y$ measures both N and the overall metallicity is immediately visible 
using the template cluster with a dispersion in [Fe/H]: $\omega$~Cen.
In Fig.~\ref{omega} the most metal-poor population defines the right edge 
of the RGB, with stars of increasing metallicity populating sequences at 
more and more negative values of $c_y$ (we used the stars studied by Johnson 
\& Pilachowski 2010 with Str\"omgren photometry by Calamida et al. 2007).
The spectral synthesis calculations reported in Sect.~\ref{synth} show
that a variation of [Fe/H] from -2.23 to -0.63, roughly the range observed
in $\omega$~Cen, causes changes in $c_y$\ of 0.11 mag for subgiants
and 0.18 mag for stars at the RGB bump, $c_y$\ being smaller in more metal-rich stars.
Such differences in $c_y$ are comparable to those expected between N-rich and N-poor
stars. Similar differences are also produced by variations in the CNO content, 
within the limits that might be expected for stars in NGC~1851 and $\omega$~Cen 
\citep[]{yg08, marinocen}. Disentangling these effects can be difficult without
further information.

\subsection{The $m_1$ index} \label{m1}

The Str\"omgren index usually employed to trace the metallicity is
$m_1$. In Fig.~\ref{figroma3} we display the effects of metallicity
using the $m_1$ index,  or better its dereddened version, \m, displaying three
representative monometallic  clusters and NGC~1851. In NGC~104 (=47 Tuc) the RGB has a
spread in \m, but this cluster is well known to have a negligible dispersion
in [Fe/H] \citep[e.g.,][]{carretta09c}. Thus, this diagram  indicates that the
index \m \ is sensible also to the abundance of elements other than iron at
high  metallicity (e.g., N is present through the CN bands visible in the $v$
filter, see  discussion on $c_y$ above and Sect.~\ref{synth}). This dependence is almost not
detectable for NGC~362,  similar in metallicity to NGC~1851, and is undetectable
in the intermediate  metallicity NGC~6752 (where the apparent spread of the RGB 
fainter than \y$\sim$13.5  is due to the blue HB stars, which form an almost
vertical sequence). 

The P and IE stars occupy the bluer and redder parts of the RGB, respectively
for  NGC~104 (and NGC 6838=M71, not shown here), while the separation is
much less clear in  NGC~362 and is not seen at all for NGC~6752 (and the other
clusters of similar or  lower metallicity). In NGC~1851 there is also a
spread, not imputable to the blue HB or simply to a metallicity difference,
since the dispersion in [Fe/H] determined from spectroscopy should yield only a
spread of 0.01 mag in the $m_1$\ index, according to our spectral synthesis (see
Sect.\ref{synth}). This cluster  is very complex and many of its feature are
discussed in \citep{carretta1851b}. 
Our second conclusion is then that \m \ {\it is a decent indicator of first and
second generation, but only at high  metallicity (indicatively for
[Fe/H]$\gsim-1.0$).}

To summarize the results of this discussion, the index $c_y$ \ is a good indicator
of different N abundances (i.e., distinct stellar generations) in metal-poor
clusters. For metal-rich clusters, like NGC~104, the variations due
to N in the $v-b$ and $u-v$ colors are similar and the net effect is to cancel
or even to change the sign of the sensitivity of $c_y$ \ to different
populations; in this metallicity regime $m_1$ \ is a better indicator. 

\begin{figure*}
\centering
\includegraphics[scale=0.95]{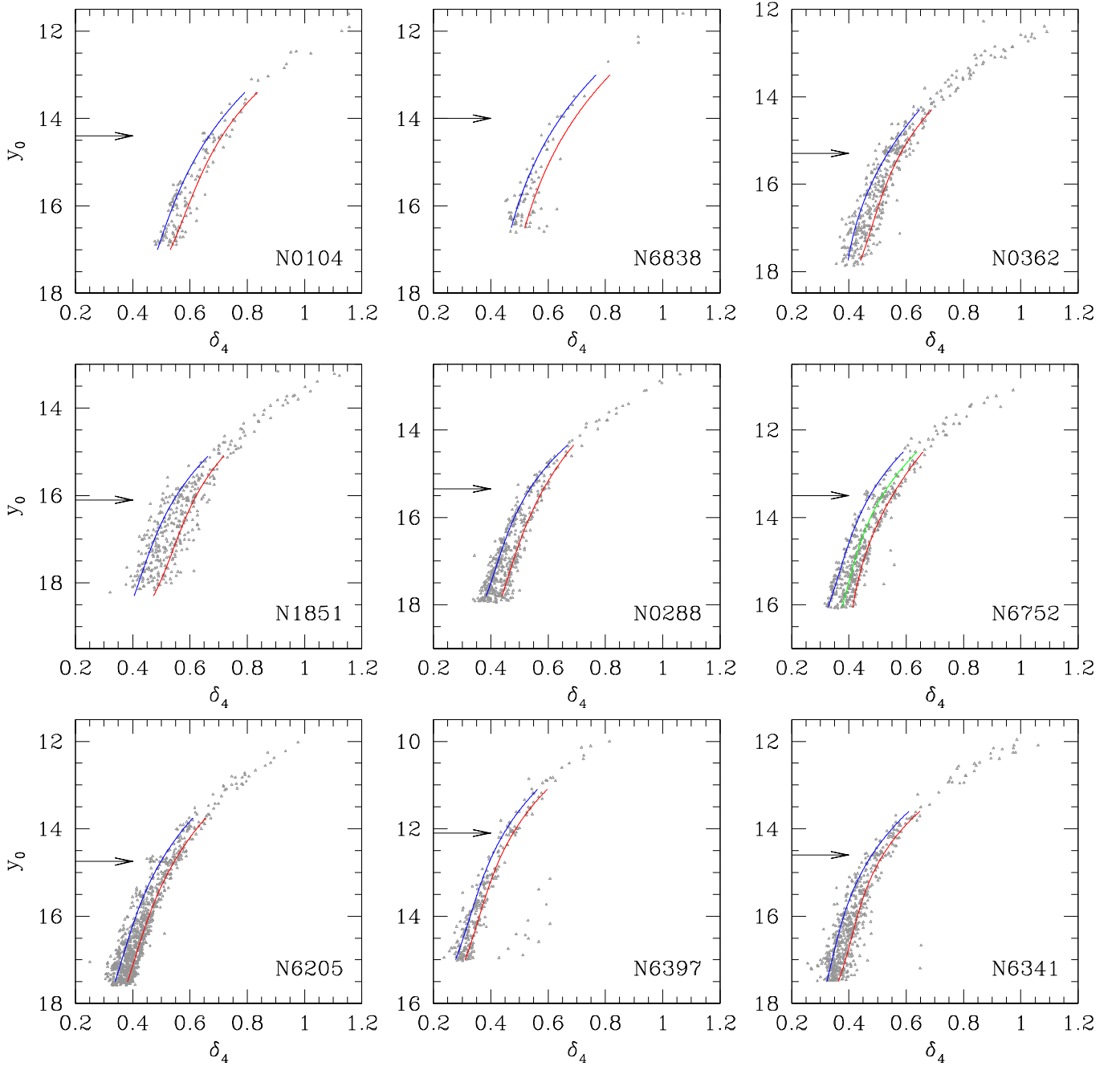}
\caption{CMDs in \del,\y \  for the RGBs of the nine GCs, with the RGB 
bump level indicated by an arrow. The blue and red curves, stopped one
magnitude brighter than the RGB bump, where the statitics begins to be too scarce, are polynomials 
describing the populations, drawn according to the distributions of P 
and IE stars in Fig.~\ref{figroma4}. For NGC~6752 the distribution of 
stars is trimodal (see text and next figure) and an intermediate (green) 
line is drawn. }
\label{delta4all}
\end{figure*}

\begin{figure*}
\centering
\includegraphics[bb=40 150 600 700,clip,scale=0.95]{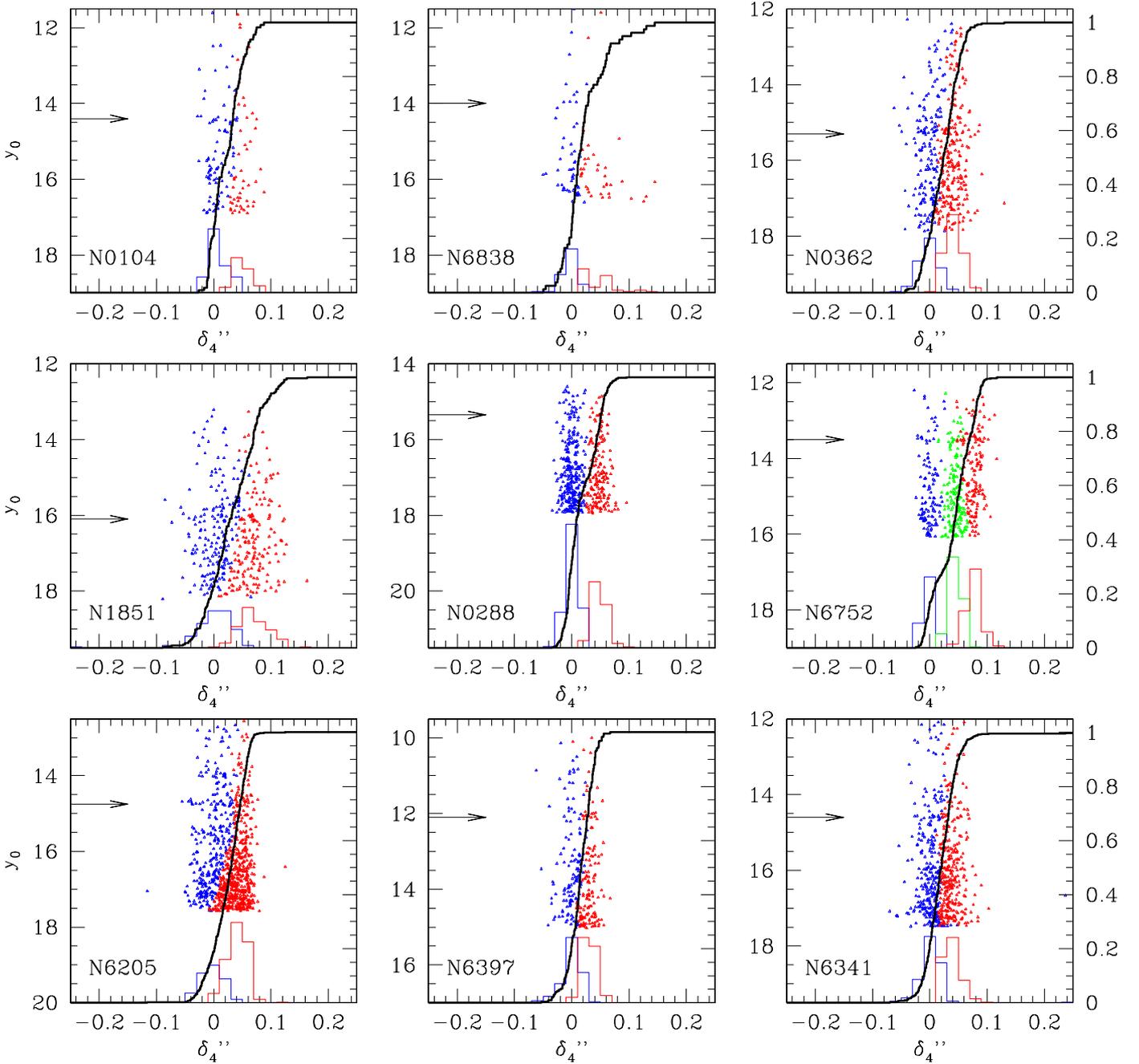}
\caption{Distribution in \del'' for the nine clusters, showing the large
difference among the GCs.  The stars are colored in blue or red (and green, 
only for NGC~6752) according to the separation made in \del ~(see previous
figure); this division is only indicative. The colored histograms refer to the
blue/red populations. The black, heavy lines represent the cumulative functions
of all stars below the RGB bump, indicated by an arrow. These cumulative
functions  (to which the left y-axes refer) are the ones to consider to
determine whether/where the distributions change slope because of a separation
in color (i.e. population). The most notable cases  are NGC~288, NGC~6752 (and
maybe NGC~104 (=47 Tuc) and NGC~362). }
\label{isto4}
\end{figure*}

\begin{figure*}
\centering
\includegraphics[bb=10 350 600 700,clip,scale=0.95]{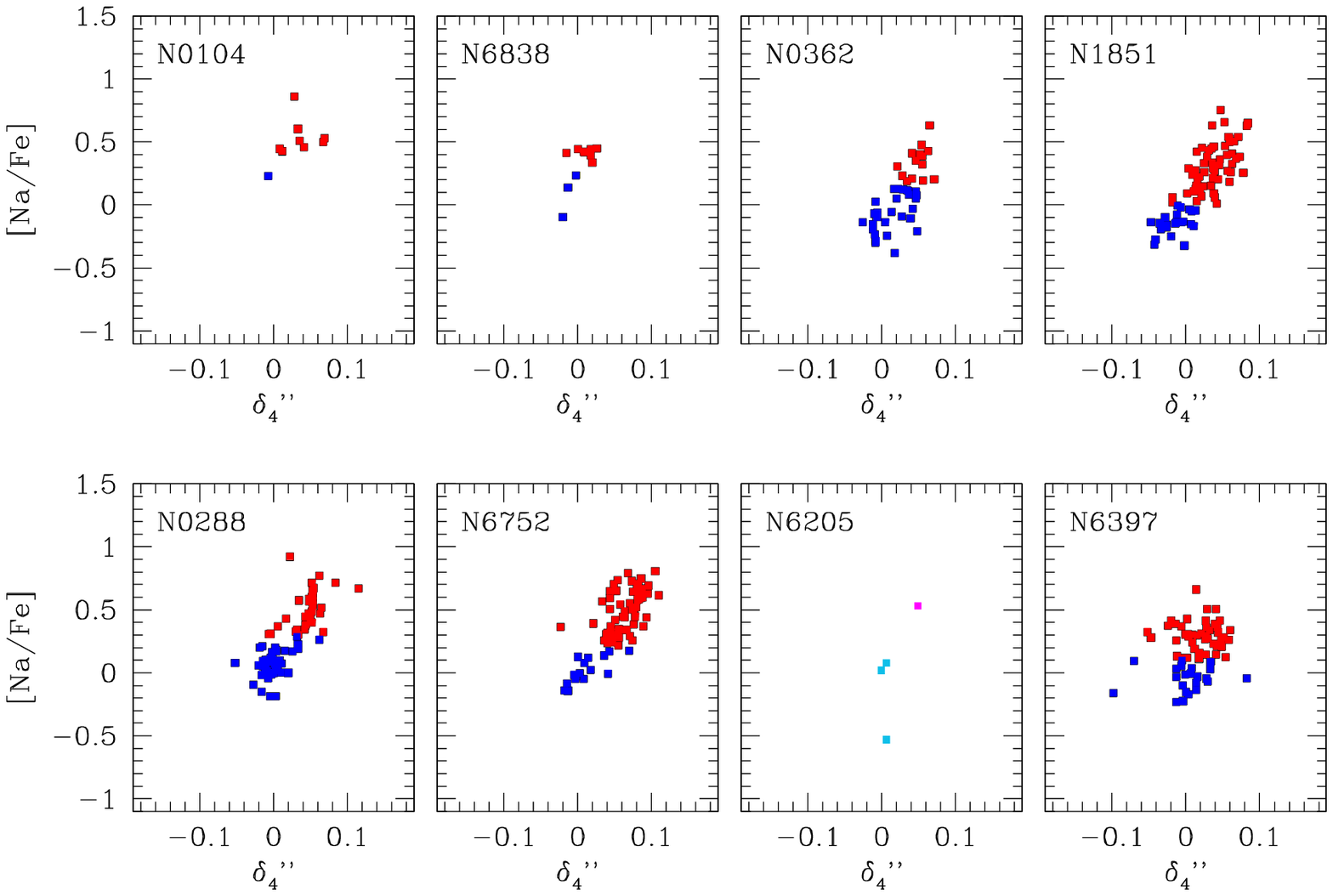}
\caption{Relation between \del'' and the Na abundance for the seven 
GCs in our spectroscopic survey (and NGC~6205=M13, shown with different colors, 
like in Fig.~\ref{figroma2}). Blue and red symbols are for P and IE 
stars, respectively (the definition is based only on Na abundance, 
to increase the sample sizes).
}
\label{nad}
\end{figure*}

\section{A new index, \del}

To reduce as much as possible any degeneracy with metallicity,  we test here a
new Str\"omgren index, that we call \del, defined as  $\delta_4=(u-v)-(b-y)$, or
$\delta_4=c_1 + m_1$. We think that it can be an  useful indicator for all
metallicities, and that it minimizes the dependence on photometric errors,
since it does not weight any of the photometric filters twice. Given the 
limit to photometric errors to $<0.02$~mag adopted to select stars, the errors
in $\delta_4$ should be $<0.08$~mag. However, a more realistic estimate is
an r.m.s of $\sim 0.02$~mag. Furthermore,
both $m_1$ and especially $c_y$ do have a stronger weight on the  bluest
filters, the ones most probably having the largest photometric errors, 
especially for red stars. In our new index the photometric errors in $(u-v)$ and
$(b-y)$ are not correlated with each other.

We have used in the present paper only dereddened values; however \del \ would
not depend much on reddening anyway. Taking from \cite{strom66} the dependence
on reddening for $c_1$ and $m_1$,  we have that $E(\delta_4)=0.02\times
E(b-y)=0.014\times E(B-V)$, i.e., \del \ is  approximatively reddening-free, at
first-order approximation.

To test the ability of \del \ to separate stars of different generations,  we
plot in Fig.~\ref{figroma4} the \del,\y \ CMDs of the nine GCs, using  the same
symbols of Fig.~\ref{figroma2}; the figure shows that there is  a good
separation  between P and IE stars. Also using this new index we see some
structure in the  RGBs, most evident for two clusters, NGC~288 and NGC~6752;
there the distribution  of stars in \del \ seems bi-(multi)modal and not
continuous. This is best shown  by Fig.~\ref{delta4all}, where we display only
the RGBs of all nine clusters on  an enlarged scale. NGC~288 appears to have two
distinct branches, one populated  by P, the other by IE stars (see the previous
figure), while NGC~6752 displays a tri-modal distribution (see also 
Fig.~\ref{isto4}). If we trust the link between \del\  and stellar generation,
we could conclude that in NGC~288 there have been two  episodes of star
formation from gas well homogenized in each phase, without any  appreciable
continuous dilution from pristine gas, as invoked by models
\citep[e.g.,][]{prantzos} to explain the run of the observed Na-O 
anticorrelations. In NGC~6752 there are three populations, even if their 
separation is not so strikingly evident as  in NGC~288.

The variation of \del \ seems to depend on magnitude, i.e. on temperature 
(Figs.~\ref{figroma4},~\ref{delta4all}) and also on metallicity, see next 
Section. Ignoring this effect would produce a blurring of the \del \ values  and
there would be the risk of potentially hiding some real features.  Guided also
by the distribution of P and IE stars that we see in  Fig.~\ref{figroma4}, we
traced reference sequences in the \del,\y \ plane. First, we fitted a cubic polynomial
in $y_0$ to the RGB ridge line; second, we labeled as ``blue" the stars with negative
residuals and ``red" those with positive residuals with respect to  it (P and IE
stars, respectively); and finally  we drew the best fitting cubic polynomial
in $y_0$ to the two distributions of blue and  red stars. The results are shown in
Fig.~\ref{delta4all} (where NGC~6752 is actually divided in three
distributions). 

These curves  can help us to eliminate the  curvature of the RGB  using both the
blue and the red sequences, and immediately  compare the distributions of red
and blue populations. We then define a new index, called $\delta_4'$  to 
rectify and normalize this index.  Of course, with this procedure we risk to 
normalize to the errors in those cases where the intrinsic spread is small with respect 
to observational errors (like e.g. the case of NGC~6397). We then defined a 
second index $\delta_4''$\ which is $\delta_4'$ multiplied by the r.m.s. scatter
around the best fit line for the whole sample (see Table~\ref{indici} for definitions). 
$\delta_4''$ takes into account the fact that the blue and red sequences can be more
or less separated in color in different clusters. This index can be used to 
better identify the presence of multiple distributions along the RGBs. Note
that given its definition, photometric errors on $\delta_4''$\ should be similar to
those in $\delta_4$, with typical r.m.s values of $\sim 0.02$~mag.

We also note that $\delta_4'$\ and $\delta_4''$ are only defined within a
limited range in magnitude for each cluster, and extrapolation of the best fit
lines out of these regimes produce results that are clearly wrong. Our following
discussion only considers those stars fainter than the RGB bump (or slightly brigther).

\subsection{\del \ and the separation of star generations}

Fig.~\ref{isto4} shows the RGBs of the nine GCs ``rectified" by using this new 
index, and allows to appreciate that strong differences do exist among the GCs
in our sample. This division is rather artificial in most cases: we simply
divided second generation (redder)  and first generation (bluer) stars by
choosing the stars that lie near the red and blue lines in Fig.~\ref{delta4all}.
The clear exceptions are NGC~288 and NGC~6752, since the distribution of giant
stars  is clearly bimodal in the first one, and tri-modal in the latter. NGC~104
(=47 Tuc)
and NGC~362 show  smaller indication of separation, although a bi-modal
distribution is not excluded; NGC~6205 (=M13) has perhaps the larger fraction of IE stars;
and NGC~1851 has the largest dispersion in $\delta_4''$, possibly in agreement with
the hypothesis of an origin  by merger of two clusters with different
chemical composition (Carretta et al. 2010b, 2011).

To effectively see whether we can really discern a separation of  populations,
in the same figure we also present the cumulative  distribution in \del''
(limited to stars below the RGB bump). A  change in the slope is an indication
of separations into discrete  distributions; NGC~104, NGC~288, and NGC~6752 are
the only three GCs where such a feature is clearly present (and maybe NGC~362).

Only in the most evident cases of NGC~288 and NGC~6752 we estimated  the
fractions of second (red, red plus green in the latter) and first  generation
(blue) stars as defined by the photometry and the \del \ (and \del'') index and
compared them with the corresponding fractions of P and IE stars,  defined 
spectroscopically. In Carretta  et al. (2009a) we found that for all GCs the
fraction of P stars is  more or less constant, at a level of about one-third.
Here we find  the same result for NGC~6752 (blue stars=27\%, red stars=73\%,
compared  to P=27\% and IE=73\%). However, for NGC~288 the fractions of P and 
IE stars are reversed (62\% and 38\% from the photometry compared to 33\% and
67\% from the spectroscopy). The split in the RGB of NGC~288  is very clear (and
is present, in the same sense, also in the \cy,\y \  CMD). There are several
considerations to be taken into account. First,  the discrepancy is reduced if
we classify stars in P and IE using only  Na abundances, since in this case we
have fractions of 42\% and 58\%,  respectively, Second, in NGC~288 the
spectroscopic sample is limited,  perhaps more than in other GCs, to stars
significantly brighter than  the bulk of the photometric sample, while the
stronger concentration of  blue/P stars is at the fainter limit of the RGB,
where the actual  separation is less evident. Third, we see in
Figs.~\ref{figroma2} and  \ref{figroma4} that, at the fainter limits of the
spectroscopic sample,  there are a few IE stars that overlap in color the P ones
(the reverse  does not happen). In this particular cluster the division in
color/index  is quite natural, while the spectroscopic classification in P and IE rests on the
less  defined minimum value for Na \citep[see][for a definition]{carretta09a}; 
indeed, a slightly different separation in Na would bring  the fractions  of P
and IE stars more in agreement with the photometric ones. In conclusion, the result
we found on the relative importance of first and second  generation stars (one
third and two thirds) is statistically valid on  the whole sample of GCs, but in
some clusters there may be individual  variations. While this issue merits
attention and further considerations,  this is besides the scope of the present
paper. 

Since the separation between populations mostly rests, spectroscopically,  on
the Na abundance, we plot in Fig.~\ref{nad} the relation between  \del'' and the
Na abundance, which is rather tight. Here we have divided  stars in P and IE
using only Na abundance,  to increase the samples (we  checked that this
approach does not generally make a significant difference  with respect to the
criteria for PIE classification as adopted in Carretta  et al. 2009a) and we
also restrict to stars below the RGB bump since the  separation between the
sequences, hence the definition of \del'', is better. We also show the case for
NGC~6205 where, however, most of the observed stars  are very bright. 
Nothing can be said for NGC~104, and NGC~6838 (=M71) since we have too 
few stars. For the other GCs, there is a very good correlation between the 
\del" index and Na, especially for the intermediate metallicity ones (the 
sensitivity is worse at very low metallicity, like for NGC~6397, where abundances 
are derived from rather weak lines). The relations could be interpreted as
continuous, but we see the possibility  of separating stars in two groups, again
most clearly in NGC~288 and NGC~6752, but also in NGC~1851 and especially in
NGC~362 (where neither \del" or Na  show well defined partitions). It is
sounder, and more natural, to separate  the populations using $both$ photometric
and spectroscopic information,  whenever possible.

\subsection{\del \ and metallicity} \label{delta4met}

Both $m_1$ and $c_y$ have a (complex) dependence on metallicity ([Fe/H] and  N
abundance). To explore what is the dependence of \del, we have  measured  the
value of \del \ at a reference magnitude ($M_V\simeq M_y=0$, i.e., near  the
level of the HB) and for the different populations, computing three  \del \
values (blue, average, and red), guided by the natural separations  and the
lines drawn in Fig.~\ref{delta4all}. Given the rather good correlation
existing between \del" index and Na abundances, we label the blue and red
lines as P and IE respectively, even though it should be clear that the
definition of PIE groups should be based on spectroscopy rather than
on photometric data. 

\begin{figure}
\centering
\includegraphics[bb=40 180 410 420,clip,scale=0.8]{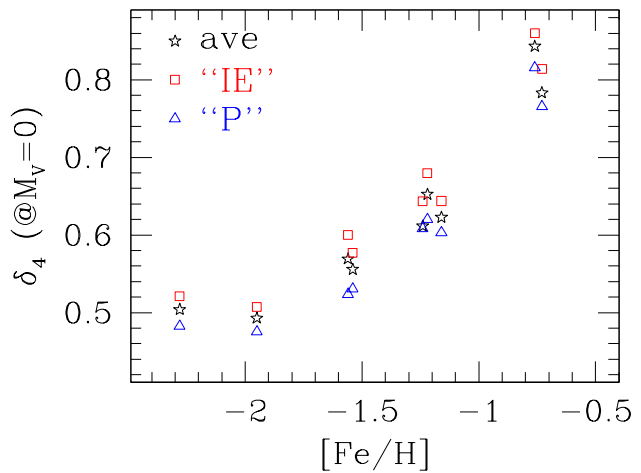}
\caption{Relation between metallicity and the value of \del \ at the 
reference magnitude $M_V=0$ (i.e., at about the HB level) computed for 
the whole RGB and for the blue and red stars (P and IE, respectively).}
\label{redde}
\end{figure}

The values we obtain by this procedure for the clusters we examined are 
plotted in Fig.~\ref{redde}. We see that \del \ has a strong dependence on metallicity.
The whole range of \del \ observed in NGC~6752 can be 
obtained if the metallicity varies by about 0.3 dex (i.e., the difference 
between NGC~6752 and NGC~288).  We then expect some difficulties in separating
N-rich and N-poor populations in clusters like NGC~1851, where there is  an
internal spread in metallicity (Yong \& Grundahl 2008 and  Carretta et al.
2010b  find a scatter of about 0.08 and 0.07 dex, respectively, well above the
expected uncertainties). We conclude that \del \ is a very good estimator of the
N abundance if a cluster is monometallic, as almost all are, at least at the
precision we can reach with the present day spectra and analysis methods (in
Carretta et al. 2009c we showed, using UVES spectra of 19 GCs, that the
internal homogeneity is at about 10\%, given an  upper limit to the scatter of
iron of less than 0.05 dex). Where this is not valid, as in NGC~1851, it is
better to separate the dependence from [Fe/H] and N using, e.g., $m_1$. Of
course,  \del \ also has a dependence on temperature, as seen from the curvature
of the RGBs.

\section{\cy, \del \ and cluster parameters} \label{param}

We have explored the possible relations of the \cy \ and \del \ indices  with
global cluster parameters. For \cy, we concentrated on the stars  below the RGB
bump where the distribution in \cy \ is almost vertical,  i.e. independent of
the temperature (Fig.~\ref{cymed}). As already  discussed, the distribution in
\cy \ is different from cluster to cluster  and may present evidence of
structures. These findings, given the connection  between this index and N, must
be indicative of different distribution in  N-richness among the GCs.

For each GC we chose only  stars in a two-magnitude bin below the RGB bump 
(i.e., $y_{bump}+0.25$ to  $y_{bump}+2.25$) and selected as RGB those  falling
within 3$\sigma$ from the average \cy \ value (after a $\kappa$-$sigma$ 
clipping); the selected region is indicated in the figure. The adopted criteria 
allow us to deal with statistically significant samples of stars and the 
verticality of the RGB in this plane allows us to separate rather easily the 
possible contamination by field stars. We measured the average value of \cy \ in
the defined interval and its r.m.s. scatter. In principle, we expect that a
higher average value does indicate a higher relative weight of the second
generation, while a larger r.m.s. scatter could be the signature of a larger
contribution of the nuclear processing by the first generation polluters to the
yields budget in the cluster.

\begin{figure*}
\centering
\includegraphics[scale=0.7]{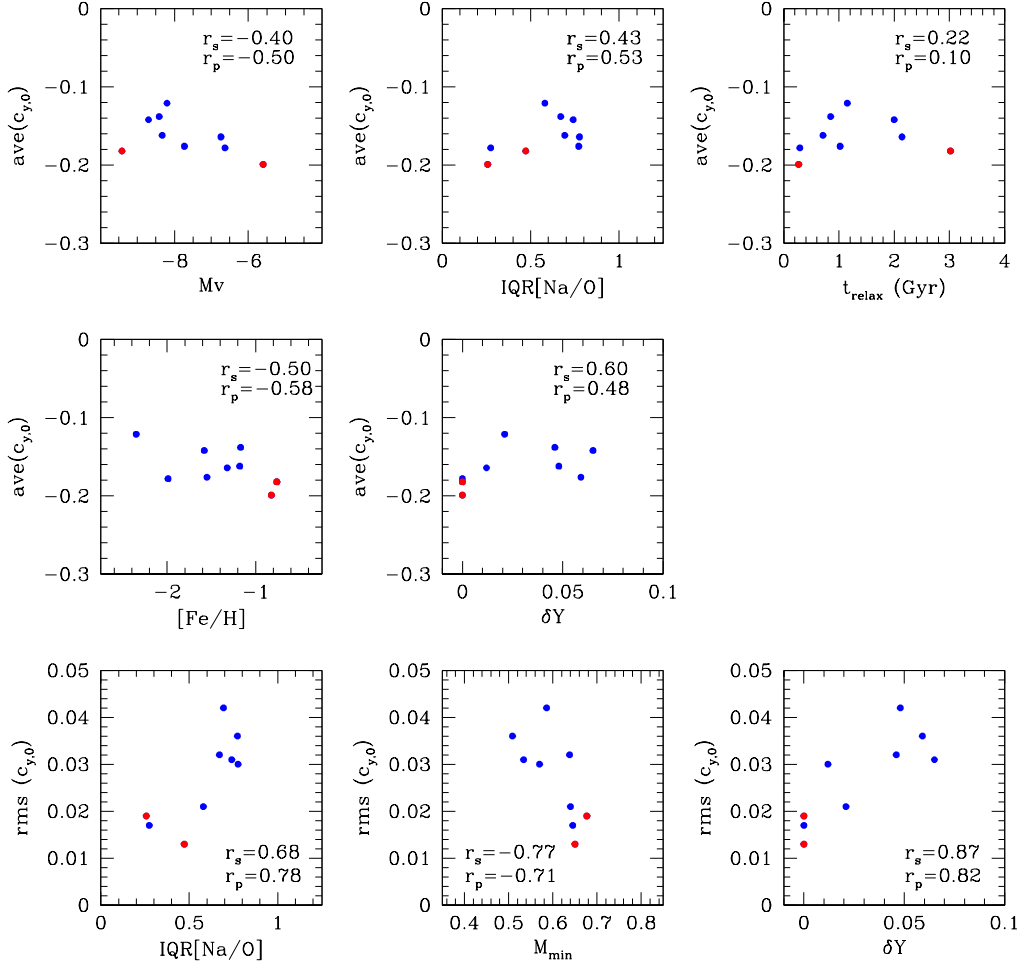}
\caption{Correlations of the average and r.m.s. for \cy \  with some interesting
parameters; the two metal-rich clusters are indicated by  red points. 
$r_S$\ and $r_P$\ are the Spearman's rank and Pearson's linear correlation
coefficients, respectively. The only highly significant correlations
(significance of the Pearson  coefficient larger than 99\%) are with metallicity
for the average value of \cy \  and with IQR([Na/O]), $M_{\rm min}$ and
$\delta$Y for its r.m.s.. }
\label{corr}
\end{figure*}

We show the  best correlations  we found for the average \cy \ and its r.m.s. in
Fig.~\ref{corr}. As expected and also already visible from  Fig.~\ref{cymed},
the average depends on metallicity (left panel in  second row of
Fig.~\ref{corr}). We have tried to see if there are other  significant
correlations with several structural or chemical properties, but results
are not very robust. The average
increases somewhat with the cluster mass (represented by its proxy,  the total absolute
visual magnitude $M_V$), and it is larger where the  Na-O anticorrelation is
more extended (as indicated by the interquartile   ratio IQR([Na/O], see
\citealt{carretta09a}). Similarly, it is related  to $\delta$Y, the amount of
He-enhancement that can explain the long  blue HB tails \citep{gratton10} and
which is connected with the extension  of the Na-O anticorrelation. Some
correlation exist with the relaxation  time at half-mass radius, once 47~Tuc
(which has the longest relaxation  time in the sample) is excluded (rightmost,
upper panel of Fig.~\ref{corr}).  A full cluster relaxation occurs in at least
2-3 relaxation times and  considered the age of 47~Tuc \citep[see,
e.g.,][]{gratton10} this GC  should be fully relaxed. 

We note that 47~Tuc was already indicated by \cite{carretta10c} and
\cite{gratton10} to be a peculiar cluster, with rather small values of 
IQR[Na/O] and He spread with respect to other clusters of similar total  mass.
The same is apparently visible in the correlation of the average  \cy \ index
and $M_V$ in the present data. Recently, \cite{lane} suggested that also this
cluster might be the result of a past merger of two originally distinct
clusters. However, we must point out that for  47~Tuc the number of stars is
smaller than in the other GCs, given the  more external region sampled by the
available Str\"omgren photometry. It  is possible that at such larger distances
from the cluster center, the  ratio first-to-second generation stars is shifted
toward the primordial  population, as expected on theoretical grounds
\citep[e.g.,][]{dercole08} and as seen e.g., from the distribution of CN-rich
stars \citep{norrisfreeman}, interpreted as Na-rich, second-generation ones in
the present-day framework.

The last row of panels in Fig.~\ref{corr} shows the best correlations we found
using the dispersion in the \cy \ index. In the case of the r.m.s. scatter,  the
correlations with the spread of He $\delta$Y (from \citealt{gratton10}) and
IQR([Na/O]) are significant at more than 99\%; the first is mirrored by the
anticorrelation between the spread in \cy and the M$_{min}$, the minimum  mass
reached along the HB (also from \citealt{gratton10}).

These findings represent  another  indication that $c_y$ is tied to the
different generations through its sensibility  to N (and the link with Na, O,
and He abundances). 

\begin{figure*}
\centering
\includegraphics[scale=0.7]{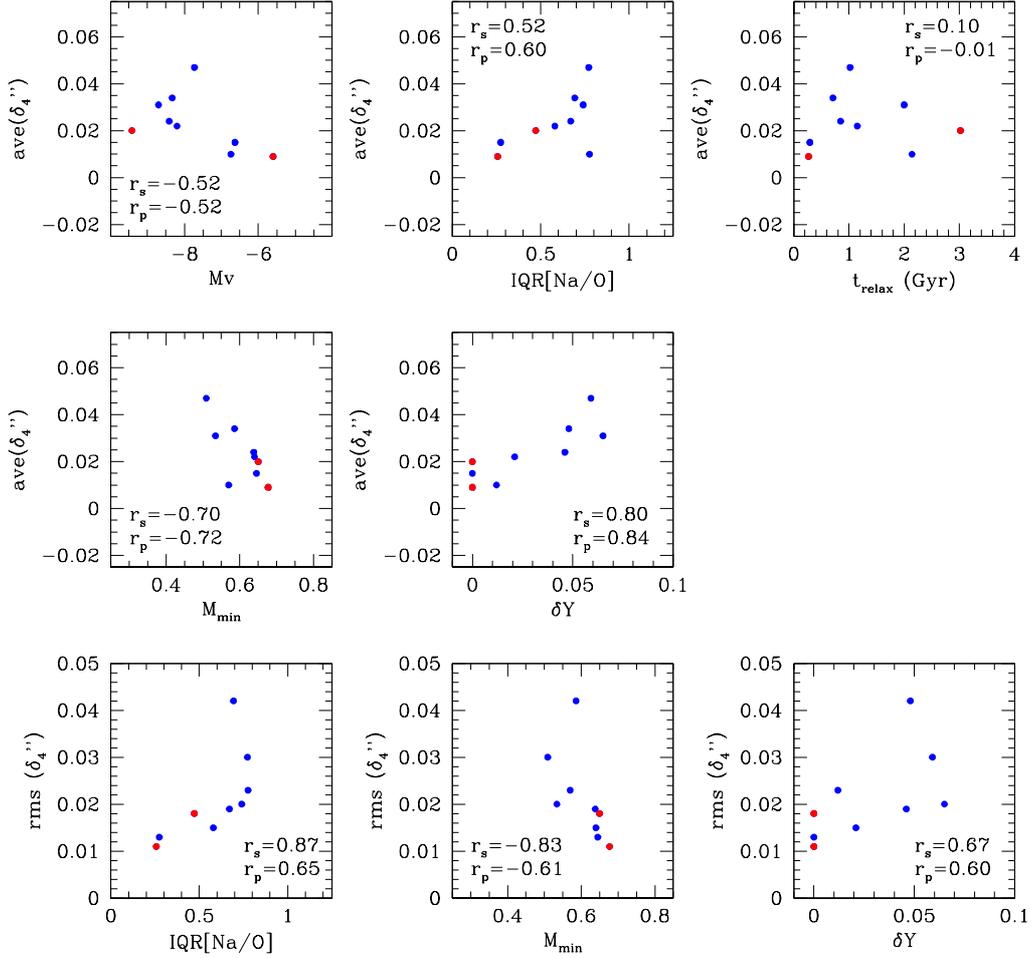}
\caption{Correlations of the median value and r.m.s. for \del \ 
with some interesting parameters. Symbols are as in previous figure.}
\label{corrdelta4}
\end{figure*}

A similar approach  can be taken for \del, and we show in Fig.~\ref{corrdelta4} 
the results. Also in this case we used stars below the RGB bump and determined 
the average and r.m.s. (we used here the median, given the less uniform 
distribution in this index) of \del" for the nine clusters. The indications  are
similar to the previous case for $M_V$,  IQR, $\delta$Y, and M$_{min}$,  while
the correlation with the relaxation time disappears.


\begin{table*}
\centering
\setlength{\tabcolsep}{1.5mm}
\caption{Model sensitivity of Str\"omgren photometric indices.}
\begin{tabular}{lcccrrrrrrrrrc}
\hline \hline
Phase &$T_{\rm eff}$ &$\log{g}$ & $[$A/H$]$ & $\Delta u$~~ & $\Delta v$~~ & $\Delta b$~~ & $\Delta y$~~ &~$\Delta Ca II$ 
& $\Delta m_1$~ & $\Delta c_1$~~ & $\Delta c_y$~~ &$\Delta \delta_4$~~ &C-normal,N-poor minus \\
\hline
SGB       &5578 & 3.44 & -2.23 & -0.056 & 0.002 & 0.000 & 0.000 & 0.003 & 0.002 &-0.061 &-0.061 &-0.059 & C-normal,N-rich \\
 RGB-bump &5018 & 2.00 & -2.23 & -0.082 & 0.003 & 0.001 & 0.000 & 0.003 & 0.001 &-0.088 &-0.089 &-0.086 & C-normal,N-rich \\
$M_V=-1$  &4857 & 1.60 & -2.23 & -0.087 & 0.001 & 0.000 & 0.000 & 0.003 & 0.001 &-0.090 &-0.090 &-0.089 & C-normal,N-rich \\
.... & .... &....  &....   & -0.062 &-0.003 & 0.000 & 0.000 & 0.003 &-0.003 &-0.056 &-0.056 &-0.059 & (a) \\					
\\
SGB       &5537 & 3.53 & -1.63 & -0.099 &-0.007 & 0.000 & 0.000 &-0.002 &-0.007 &-0.086 &-0.086 &-0.092 & C-normal,N-rich \\
 RGB-bump &5001 & 2.19 & -1.63 & -0.144 &-0.043 & 0.000 & 0.000 &-0.005 &-0.043 &-0.059 &-0.059 &-0.102 & C-normal,N-rich \\
$M_V=-1$  &4683 & 1.51 & -1.63 & -0.175 &-0.102 &-0.002 & 0.000 &-0.007 &-0.101 & 0.027 & 0.028 &-0.071 & C-normal,N-rich \\
.... &.... & .... & ....  & -0.096 &-0.043 & 0.000 & 0.000 &-0.005 &-0.043 &-0.009 &-0.009 &-0.052 & (a) \\
\\
SGB       &5369 & 3.44 & -1.23 & -0.138 &-0.022 & 0.001 & 0.001 & 0.000 &-0.023 &-0.093 &-0.093 &-0.116 & C-normal,N-rich \\
 RGB-bump &4948 & 2.42 & -1.23 & -0.176 &-0.076 & 0.001 & 0.000 &-0.002 &-0.078 &-0.023 &-0.024 &-0.101 & C-normal,N-rich \\
.... &.... &....  &....   & -0.221 &-0.163 &-0.010 &-0.002 &-0.012 &-0.161 & 0.096 & 0.103 &-0.050 & C-rich,N-rich \\
$M_V=-1$  &4478 & 1.40 & -1.23 & -0.189 &-0.149 &-0.002 & 0.000 &-0.006 &-0.145 & 0.107 & 0.109 &-0.038 & C-normal,N-rich \\
.... &.... &....  &....   & -0.014 &-0.025 &-0.006 &-0.002 &-0.021 &-0.015 & 0.030 & 0.034 & 0.015 & C-rich,N-poor \\
.... &.... &....  &....   & -0.251 &-0.274 &-0.027 &-0.006 &-0.021 &-0.226 & 0.270 & 0.291 & 0.044 & C-rich,N-rich \\
.... &.... &....  &....   & -0.162 &-0.103 &-0.001 & 0.000 &-0.006 &-0.101 & 0.042 & 0.043 &-0.058 & (a) \\
\\
SGB       &5229 & 3.71 & -0.63 & -0.190 &-0.110 &-0.001 & 0.000 &-0.003 &-0.109 & 0.029 & 0.030 &-0.080 & C-normal,N-rich \\
 RGB-bump &4765 & 2.53 & -0.63 & -0.211 &-0.180 &-0.003 & 0.000 &-0.007 &-0.175 & 0.146 & 0.149 &-0.029 & C-normal,N-rich \\
$M_V=-1$  &4016 & 1.08 & -0.63 & -0.153 &-0.176 &-0.005 &-0.001 &-0.011 &-0.166 & 0.193 & 0.197 & 0.027 & C-normal,N-rich \\
.... &.... &....  &....   & -0.097 &-0.123 &-0.002 & 0.000 &-0.011 &-0.118 & 0.146 & 0.148 & 0.028 & (a) \\
\hline		       
\end{tabular}
\tablefoot{(a) $\Delta$ computed as C-normal,N-poor minus C-normal,N-rich, but with a composition changed after the RGB bump for C,N.
\\
C-normal,N-poor stars have: [C/Fe]=[N/Fe]=0, [O/Fe]=[Mg/Fe]=[Ca/Fe]=+0.4, 
[Ti/Fe]=0.2; \\
C-normal,N-rich stars have [C/Fe]=$-0.2$, [N/Fe]=+1.3, [O/Fe]=$-0.1$ (and the same abundances for Mg, Ca, and Ti); \\
C-rich,N-poor stars have [C/Fe]=+0.2, [N/Fe]=0.0, O and other elements as in first case;\\
C-rich,N-rich stars have [C/Fe]=+0.2, [N/Fe]=+1.3, [O/Fe]=$-0.1$ and other elements as in the first case.}
\label{t:synt}	       
\end{table*}

\section{Comparison of observed colors with synthetic spectra: interpretation}
\label{synth}

In order to better understand the observed runs of the classical and new indices
we obtained from the Str\"omgren system, we computed a number of  synthetic
spectra over the wavelength region of interest (approximately, 
$\lambda=3000-6000$~\AA) for some cases of interest. The synthetic spectra were
computed using the \cite{kur93}  set of model atmospheres (with the overshooting
option switched off), and line lists from \cite{kur93} CD-ROM's. 
It is well known that this line lists, while very extensive and useful, still
have important limitations. 
In addition, both the model atmospheres and our syntheses include 
approximations (e.g. mono-dimensional atmospheres, scattering treated as an
additional absorption, etc.). For this reason we only  consider differential
effects between stars that should be very similar, except for details of
the chemical composition. 

All synthetic spectra were computed with a wavelength step of 0.02~\AA. Then
they were convolved with a Gaussian with a FWHM of 2~\AA, and rebinned at about 
1~\AA \ sampling for an easier handling and display. The spectra were
finally convolved with the transmission of the $uvby$\ \citep{strom56} and $Ca$\
filters \citep{hk}, collected in the Asiago Database on Photometric Systems 
\citep{asiago}.

For all choices of atmospheric parameters, we considered as reference the
spectra computed with the typical abundance pattern observed in field 
metal-poor stars, that is [C/Fe]=[N/Fe]=0, [O/Fe]=[Mg/Fe]=[Ca/Fe]=0.4, and
[Ti/Fe]=0.2. Spectra computed with these parameters are labeled ``N-poor". They
should mimic the composition of P stars according to \cite{carretta09a}. For the
same choices of atmospheric parameters, we computed spectra for atmospheres
having [C/Fe]=$-0.2$, [N/Fe]=+1.3, and [O/Fe]=$-0.1$ (and the same abundances
for Mg, Ca, and Ti). We labeled this second group of spectra ``N-rich"; they
should mimic the spectra of I stars.\footnote{We do not expect any significant 
change between I and E stars in this respect, insofar the abundance of
O is larger than that of C. The main difference between I 
and E stars is in the O abundance, which changes from [O/Fe] about $-0.1$ to 
about $-1$, but even in this case spectroscopic observations do not
indicate the presence of very strong C-bands, which are expected as soon
as the abundance of C exceeds that of O. The filters we are considering do 
not contain OH bands (e.g., the 
strong ones near 2700 \AA) but only N molecules. However, N is already about 90\%
of the whole sum of CNO elements in I stars, and the small further increase 
expected between I and E stars (about 5\%) does not influence our computations. } 
In addition, we also explored in a more limited 
way the influence of C increase, computing spectra for [C/Fe]=0.2, [N/Fe]=0.0, 
[O/Fe]=0.4 (``C-rich,N-poor" spectrum) and [C/Fe]=0.2, [N/Fe]=1.3, [O/Fe]=-0.1 
(``C-rich,N-rich" spectrum). All these compositions are also indicated in 
Table~\ref{t:synt}.

\begin{figure}
\centering
\includegraphics[bb=50 170 525 700,clip,scale=0.55]{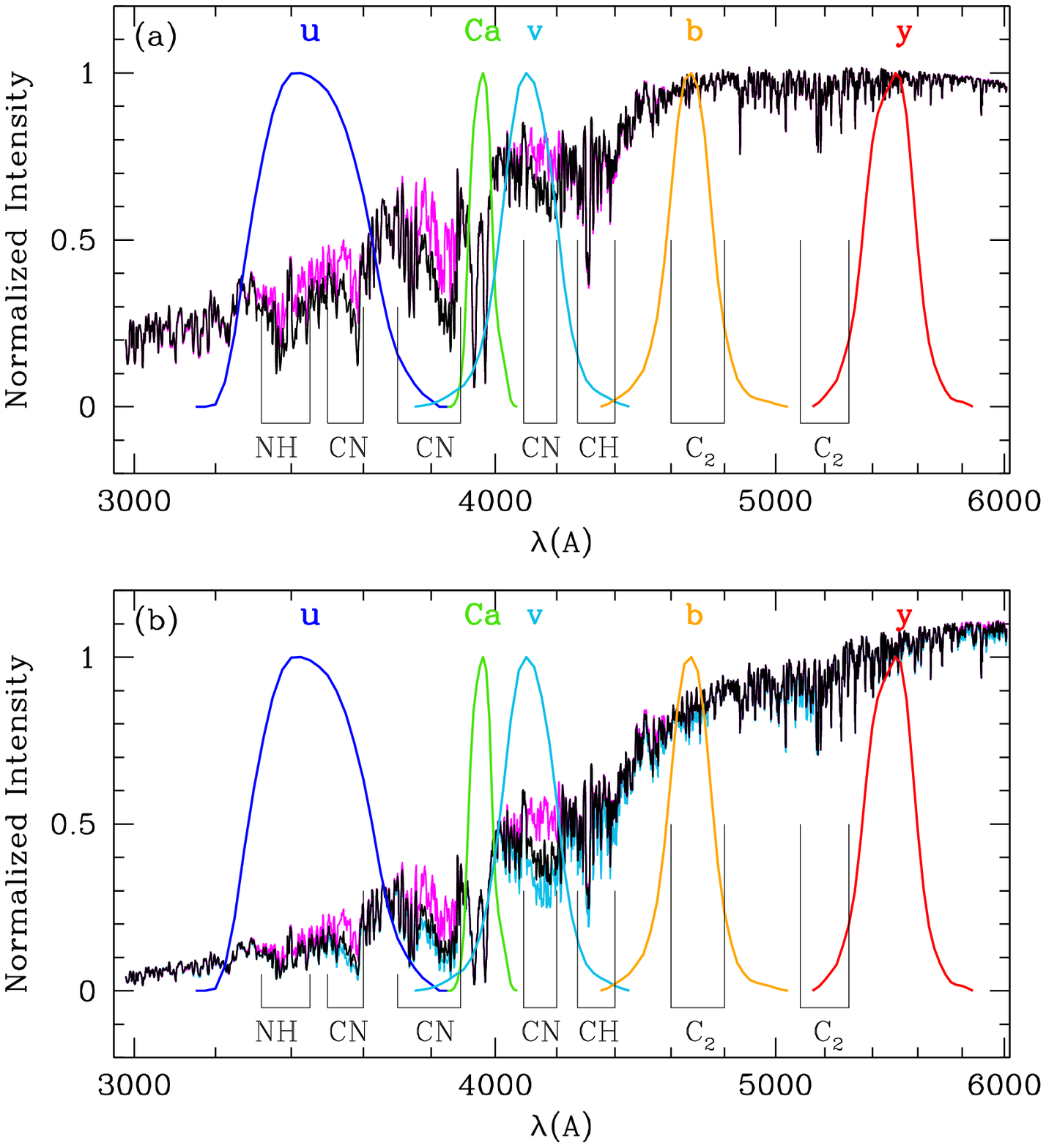}    
\caption{(a) Synthetic spectra for a N-poor
(magenta), a N-rich (black), and a C-rich star at the RGB bump of a GC with
[Fe/H]=-1.23 (see Sect. 4 for the various abundances used). Over-imposed are 
the transmission curves for the $uvbyCa$ filters; some features of interest 
are indicated (see text).  (b) The same, but for a star above the RGB bump.}
\label{f:band}
\end{figure}

First we considered  pairs of ``N-poor" and ``N-rich" spectra for the four
metallicities [Fe/H]=$-2.23$, $-1.63$, $-1.23$, and $-0.63$. They have
effective temperature $T_{\rm eff}$\ and surface gravity 
$\log{g}$\ appropriate for stars at the base of the subgiant branch (SGB), at 
the RGB bump, and at $M_V=-1$ (as read from the isochrones of 
$\log$(age)=10.10 by \citealt{bertelli}). Table~\ref{t:synt} lists
the values of the atmospheric parameters we used and the
offset in magnitude between the ``N-poor" and ``N-rich" spectra
for the $uvby$\ and $Ca$\ bands and for various photometric
indices ($m_1$, $c_1$, $c_y$, and $\delta_4$). 

\begin{figure*}
\includegraphics[bb=10 340 565 700,clip,scale=0.9]{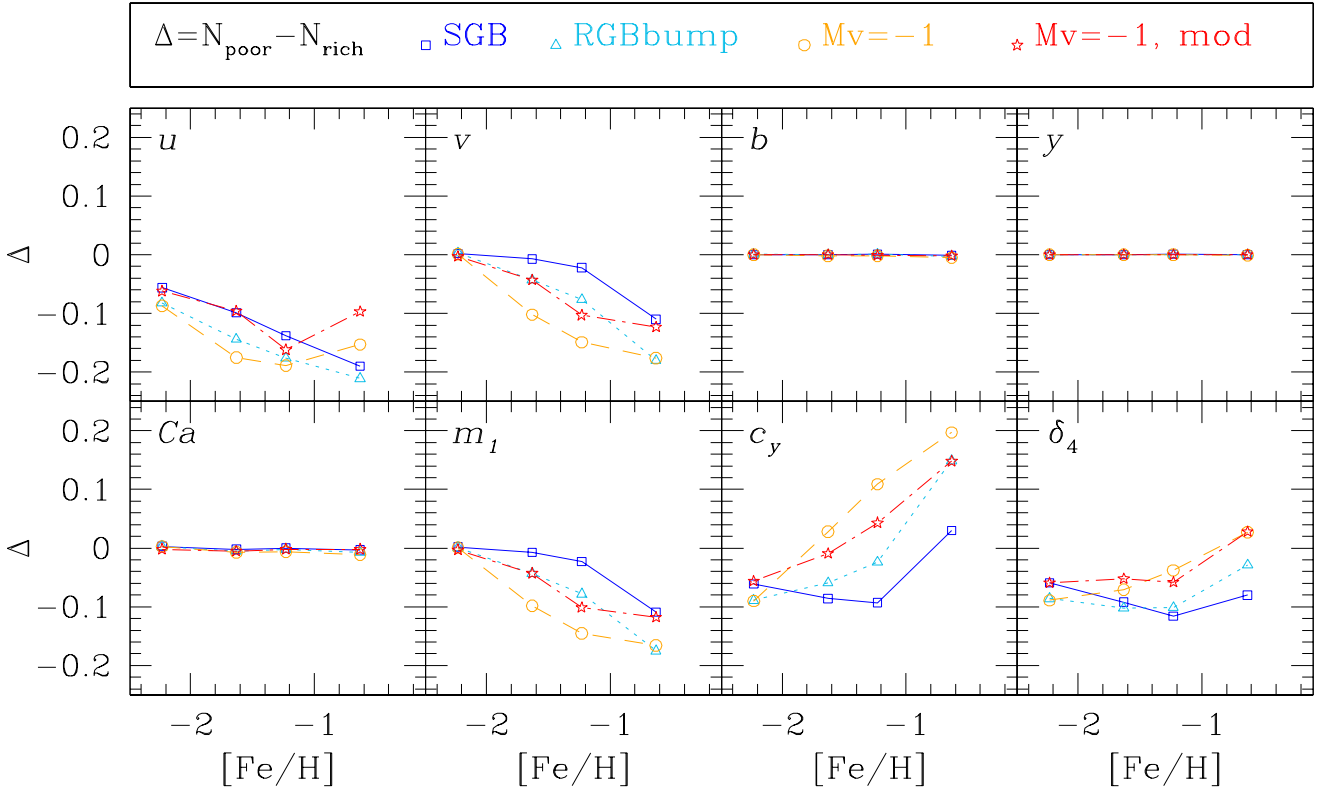}
\caption{Run with metallicity of the difference in magnitude and color for 
``N-rich" and ``N-poor" stars (see Table~\ref{t:synt}). We show the three
different evolutionary phases considered (two different compositions for the 
brighter star, see text).  In each panel we indicate the filter/index involved.} 
\label{deltacol}
\end{figure*}

Figure~\ref{f:band} presents an example for the synthetic spectra appropriate to
stars at the RGB bump (panel a) and above the RGB bump (panel b) for [Fe/H]=$-1.23$;  
the transmission of the filters is also shown.
This figure visually demonstrates what has already been discussed in the previous
sections. The differences between the ``N-poor"  and ``N-rich" spectra are
essentially due to the different strength of the molecular bands: the NH band at
$\sim3400$~\AA, which is stronger  in the ``N-rich" spectra and falls within
the $u$ band; and the CN violet band at $\sim4216$~\AA, which falls within the
$v$ band. The CH A-X band at $\sim4320$~\AA\ is also different between the two
spectra, but is near the edge of the $v$\ band, so that its impact on the
Str\"omgren colors is very small. As a consequence, the flux predicted for both
the $v$\ and the $u$\ bands is smaller for the ``N-rich" spectra than for the
reference ``N-poor" ones. The difference is larger for the $u$ band, where it
may be as much as 0.2 mag, save for the coolest and more metal-rich stars, where
the effect is larger in the $v$\ band. This is due to both saturation and
different temperature sensitivity of the molecular bands. A sketch of the 
variations in the $u,v,b,y,Ca$ filters and in the three indices $m_1$, $c_y$, 
and \del \ is presented in Fig.~\ref{deltacol} where we can appreciate their 
different sensitivity to CNO abundances along the various evolutionary phases 
and their run with metallicity. Table~\ref{t:synt} and Fig.~\ref{deltacol} can 
be used to interpret what we have seen in the observed CMDs in term of 
different chemical composition (in other words, of different stellar population).

While the largest variations between first and second generation stars are 
expected for N, also the C variations can have an effect on the 
photometry. We computed only a couple of spectra, and Fig.~\ref{f:band} 
shows also the results a for C-enhanced composition. The corresponding 
differences in the various filters/indices are indicated in Table~\ref{t:synt}. 
From these computations we see that, for N-poor stars the effect of an 
increased C abundance is small, since the CN bands are not very strong 
and O is much more abundant than C in any case. On the contrary, for 
N-rich stars, an enhancement in C produces large changes to the Str\"omgren 
magnitudes, especially for the $u$ and $v$ filters. This means large 
differences in indices like $c_y$ and $m_1$ (while the differences almost 
cancel out for \del, since it is the sum of the previous indices, whose 
variations have opposite sign).

The combination of N and C variations could then explain some peculiar 
cases, like NGC~1851, where we see cluster stars lying along an ``anomalous", 
less populated RGB (see e.g. Fig.~\ref{figroma3}). In this cluster there 
have been (controversial) claims of variations of the sum  of C+N+O; this 
will be discussed in a paper in preparation. Here we simply propose that 
these stars could be the ones which are both C and N-rich, while the ones 
which have lower N abundance fall in with the bulk of the stars, even if C-rich.

Finally, we considered another possible composition difference; at the 
RGB bump there is a change in the surface abundance of C and N 
\citep[see e.g.,][]{gratton00}, i.e., C decreases and N increases, due to 
a mixing episode. In C-normal, N-poor stars the increase in N after the 
RGB bump almost compensates the decrease in C, and the CN bands do not 
change their strength. Instead, in C-normal, N-rich stars, the N abundance 
is already high and the N increase has proportionally a smaller impact and 
does not compensate enough the decrease in C. Hence, there is a smaller 
difference between a N-rich and N-poor star if we take into account the 
evolutionary changes (see the correspondingly smaller differences in $u$ 
and $v$ and all correlated indices).

Armed with these results, we may then examine the prediction of our
spectral synthesis on the observed indices. 

$m_1=(v-b)-(b-y)$ is the classical metallicity index of the Str\"omgren
photometry. In the present context, we notice that $b$ and $y$  (hence $b-y$,
the classical temperature indicator) are insensitive to variations in the 
abundances of the CNO elements. For $m_1$, the only dependence is then contained
in the $v$\ magnitudes. Since the CN violet band is very weak in  metal-poor
and/or warm stars, $m_1$\ separates ``N-poor" and ``N-rich" only for cool,
metal-rich stars.

$c_1=(u-v)-(v-b)$ is the classical gravity index of Str\"omgren
photometry, measuring the Balmer jump. In the present context, it
essentially measures the difference between the strength of the NH and
CN violet bands (the latter actually weighted twice). Therefore it 
separates well ``N-poor" and ``N-rich" at low metallicities (where the CN 
violet band is negligible), but it changes the sign of its sensitivity
for cool metal-rich stars, where the CN violet band is quite strong. The 
same holds for $c_y$ ($=c_1-(b-y)$), which has the same sensitivity on CNO
abundances of $c_1$.

Our new index $\delta_4=(u-v)-(b-y)$\ keeps constant the sign of its 
sensitivity to CNO abundances over a range of temperature/metallicity/gravity 
larger than for $m_1$ and $c_1$ or $c_y$, although it also
fails for the coolest and most metal-rich stars. The reason of this more
uniform behavior is that it weighs only once the $v$\ band, and it is then
less sensitive to the strength of the CN violet bands. In practice, the 
sensitivity of $\delta_4$\ on CNO abundances is all contained in the $(u-v)$\ 
term. The addition of the $(b-y)$ term is however useful to reduce its 
sensitivity to temperature and reddening. In particular, since $\delta_4$\ is 
practically insensitive to reddening, it can be safely used also for clusters 
with strong differential reddening.

\subsection{Comparison with Sbordone et al. (2011)} \label{sbordone}

This very recent paper approached the problem of the photometric signatures
of multiple populations in GCs from the theoretical point of view. The
approach and purposes were different, since the authors wanted
to present there full blown isochrones and take the effect of helium in
consideration as well,while we concentrated on the RGB. They then 
computed self-consistent stellar models, taking into account the different 
chemical composition of first and second generations stars both for the 
stellar interiors and the stellar atmospheres. They computed models with 
several composition (standard mixture, NaCNO-enhanced, He-enhanced) and 
translated the theoretical isochrones into the observational planes for 
Johnson and Str\"omgren photometry, showing the effects on the main sequence, 
SGB, and RGB.

Our paper and theirs, while addressing the same problem, differ in several
points: (i) while we have worked on the observational data of several GCs, they
did not  present a direct  comparison with observed cases, deferring this to
subsequent  papers; (ii) we have not computed new, self-consistent atmospheres
and fluxes,  but only computed the variation in the absorption spectrum due to a
change in  composition, using a standard model. Their approach is formally
better; however  the effect is small, as shown also by their paper, since the
molecular bands  present in these stars affect only a small fraction of the
total flux and the  various models differ only in the most external parts, near
the stellar surface.  In fact, \cite{gustafsson} considered the impact of mild
and heavy CN cycling on  the structure of the atmospheres and found that it is
negligible for the  temperatures of interest in our computations (see their
Fig.~7); (iii) in our paper  the changes in composition due to the first
dredge-up and the extra-mixing after  the RGB bump are taken into consideration.
This has significant impact on the  derived results, since C and N are involved
in these changes; (iv) perhaps most  importantly, they do all calculations only
for one value of metallicity, while we  explore {\em the entire range of metallicity
of Galactic GCs}. This is an important  factor, since we have seen that there is
a strong dependence on metallicity of  the sensitivity of the various colors and
indices to different compositions  (see Table~\ref{t:synt} and
Fig.~\ref{deltacol}).

Given all these differences, it is legitimate to ask whether we are obtaining
similar results. Taking our case  at [Fe/H]=$-1.63$ to compare to their $-1.62$
and looking at their Fig. 8, we  obtain similar, but smaller, differences
between N-rich and poor compositions  in $u$ (and of course $b$ and $y$, where N
has no influence). There are instead discrepancies in the $v$ filter, where we
find differences and they do not. However, there is a large difference in C
abundance between the two cases: we  use [C/Fe]=$-0.2$, compared to their
$-0.6$, meaning that their model has  [C/H]$=-2.23$, i.e., is more similar to
our [Fe/H]$=-2.23$ case, where we too  do not see any effect in the $v$ filter
(the CN bands are really very weak in  these models).

The large difference in C abundance and the single, quite low metallicity could 
also explain the fact that their models predict larger differences  in $c_y$
between the different compositions (see their Fig.~15) than our  computations
(Table~\ref{t:synt}). In practice, we have stronger CN bands in our  spectra and
this has an effect on the $v$ filter, that is actually weighted twice  in the
definition of $c_y$, and outweighs the effect in the $u$ filter, where  the NH
band dominates. A more complete comparison between the two works is  difficult
and outside the goals of the present work. It could be interesting,  however, to
merge the two approaches, to better define the properties of multiple 
populations in GCs.

\section{Summary}\label{conclu}

In this paper we explored the capabilities of the Str\"omgren photometric system
to characterize multiple populations that we firmly know (from spectroscopy) to
be harbored in {\em all} GCs. Thanks in particular to the $u$ filter, which
encompasses the molecular NH band at 3400~\AA, and the $v$ filter, at least  for
the metal-rich GCs, it is possible to well separate N-poor (Na-poor, O-rich,
first generation) stars from N-rich (Na-rich, O-poor, second generation) stars
using Str\"omgren multicolor photometry.

In our systematic exploration of the parameter space, using Grundahl's (and
coworkers) photometry made available by Calamida et al. (2007) for 9 GCs, we 
found that the classical metallicity index $m_1$ is able, in the present
context, to separate in N-strength classes only cool and metal-rich stars. 

Both the typical gravity indicator $c_1$ and its improved version $c_y$ defined 
by Yong et al. (2008), work well in separating N-rich and N-poor stars at low
metallicity, but give ambiguous results in the metal-rich and cool stars regime.

We defined a new indicator, $\delta_4$, that seems to works better than the
previous indices over a larger range of metallicity, although even in this case
the separation of different populations is not achieved well in the cool  and
metal-rich regime.

The variations that are sampled by all these indices are well accounted for by
changes in the abundances of CNO elements, as demonstrated by a number of
comparison with synthetic spectra convolved with the transmission filter.
However, further work is necessary, combining for instance our observational 
approach with the theoretical one chosen by \cite{sbordone}.

In the present paper we have limited our analysis to the RGB stars to ensure 
the best photometric data and to compare directly with the available firm 
constraints given by the chemical abundances (e.g., the Na-O anticorrelation).
The next step will be the extensive analysis of photometric data over wide
fields and reaching also extremely blue (faint) HB and main sequence stars  with
comparable precision. The gain in statistics will permit to better study  hot
issues such as e.g., the radial distribution of multiple stellar  populations in
GCs, which is tightly connected with the formation and evolution  of these
stellar aggregates.

\begin{acknowledgements}
We wish to thank the referee Luca Sbordone for his careful report, certainly
useful to improve the paper.
We used data from the Two Micron All Sky Survey, which is a joint project of 
the University of Massachusetts and the Infrared Processing and Analysis 
Center/California Institute of Technology, funded by the National Aeronautics 
and Space Administration and the National Science Foundation. This research 
has made use of the SIMBAD and ViZier databases, operated at CDS, Strasbourg, 
France and of NASA's Astrophysical Data System. Financial support from 
PRIN-MIUR 2007 ``Multiple Stellar Populations in Globular Clusters:
Census, Characterization and Origin", and PRIN-INAF 2009 Formation and Early 
Evolution of Massive Star Clusters", is acknowledged.
\end{acknowledgements}

\end{document}